%% file: main.tex
\documentclass {thesis}
\usepackage{parcolumns}

\input{tex/math}
\date{May 31, 2021}
\makeatletter
\def\subtitle#1{\gdef\@subtitle{#1}}%
\def\@subtitle{\@latex@error{No \noexpand\subtitle given}\@ehc}%
\makeatother

\title{\vspace{-3cm}Designing and developing tools to automatically identify parallelism}
\author{Fabian Mora Cordero}
\subtitle{Prelim committee: \\
Advisor: Dr. Sunita Chandrasekaran \\
Dr. Stephen Siegel\\
Department of Computer \& Information Sciences \\
University of Delaware
}

\makeatletter
\def\@maketitle{%
  \newpage
  \null
  \vskip 1em%
  \begin{center}%
  \let \footnote \thanks
    {\LARGE \@title \par}%
    \vskip 1.5em%
    {\large
      \lineskip .5em%
      \begin{tabular}[t]{c}%
        \@author
      \end{tabular}\par}%
    \vskip 1em%
    {\large \@date}%
    \vskip 0.75em%
    {\large \@subtitle\par}%
    \vskip 1.5em%
  \end{center}%
  \par
  \vskip 1.5em}
\makeatother

\begin{document}
\maketitle

\begin{abstract}
    In this work we present a dynamic analysis tool for analyzing regions of code and how those regions depend between each other via data dependencies encountered during the execution of the program. We also present an abstract method to analyze and study parallelism in a directed graph, by studying a Quotient Graph of the execution graph of a program, and give a simple algorithm for searching parallelism in execution graphs with a high degree of symmetry. Finally, we evaluate our approach selecting four dwarfs out of 13 Berkeley’s computational dwarfs or otherwise known as parallel patterns. 
\end{abstract}

\input{tex/1_introduction}

\input{tex/2_tool}
\input{tex/3_execution}
\input{tex/4_results}
\clearpage 
\input{tex/5_conclusions}

\clearpage 
\printbibliography
\end{document}

%% file: tex/math.tex
\newcommand{\abs}[2][]{\vert #2 \vert_{#1}}

\DeclareMathOperator*{\Aut}{Aut}
\DeclareMathOperator*{\Equiv}{Equiv}
\DeclareMathOperator*{\ExecT}{ET}

\newcommand{\Path}{\rightsquigarrow}
\newcommand{\Edge}{\rightarrow}

\usepackage{listings}
\usepackage[end]{algpseudocode}
\usepackage{algorithm}
\usepackage{algorithmicx}

\definecolor{light-gray}{gray}{0.95}

%% file: tex/1_introduction.tex
\section{Introduction}
In the last couple of decades parallel computing has garnered mainstream interest from computer scientists and in general computer programmers alike. This happened because computer processor manufacturers changed chip designing practices opting for pipelined or superscalar architectures and chip parallelism instead of increasing clock frequencies like they did until 2005, which happened for a variety of reasons, among them power and heat dissipation issues \cite{cpu-freq-web}.

There has been a paradigm shift from homogeneous systems to heterogeneous systems over the past decade or two. The heterogeneous system era has seen the introduction of massively parallel architectures such as GPGPUs from vendors such as NVIDIA and AMD, vector architectures such as the A64fx from Fujitsu/arm and also other types of devices such as FPGAs, ASICs, neural engines, many core processors and so on. Architectures are evolving continuously leading to a perpetual disruption in software. Developers are facing a non-trivial challenge porting existing algorithms, creating newer parallel algorithms where need be, parallel friendly data structures and  new programming models that offers the features to express parallelism in a thorough yet easy manner~\cite{gerber2018crosscut}. 

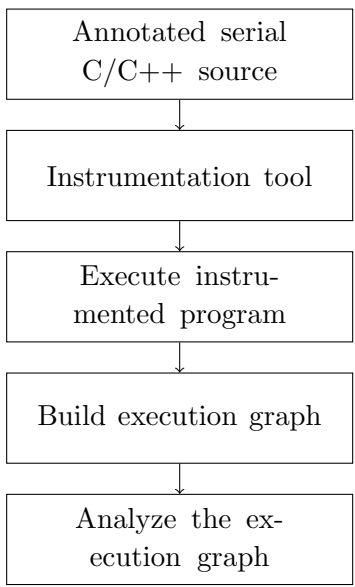
\begin{figure}
    \centering
    \begin{tikzpicture}[node distance = 4mm and 12mm, rect/.style = {rectangle, draw, align=center, text width=45mm, minimum width=4mm, minimum height=12mm, inner sep=0.5mm}]
        \node[rect] (A) {\small Annotated serial C/C++ source};
        \node[rect, below=of A] (B) {\small Instrumentation tool};
        \node[rect, below=of B] (C) {\small Execute instrumented program};
        \node[rect, below=of C] (D) {\small Build execution graph};
        \node[rect, below=of D] (E) {\small Analyze the execution graph};
        \draw[->] (A) -- (B);
        \draw[->] (B) -- (C);
        \draw[->] (C) -- (D);
        \draw[->] (D) -- (E);
    \end{tikzpicture}
    \caption{Workflow of the proposed method.}
    \label{fig:workflow}
\end{figure}

In this work we present a dynamic analysis tool for C/C++ sequential programs for exploring available parallelism within the execution of a program. The workflow of the proposed method is presented in Figure \ref{fig:workflow}, starting with an annotated C/C++ source code and returning an analysis of the execution graph of the program.

The proposed method is based on a task programming model. In this model data dependencies create interrelations between executed tasks and tasks to be executed during a program execution, producing what we call the execution graph of a program. This graph will then be used to explore parallelism in the program. 

The proposed method in its current state has the limitation that the results provided by the analysis of the execution graph are directly tied to a specific set of input parameters, thus they can not be automatically generalized to any input parameters, however we believe that they provide an insightful view of the parallelization opportunities that the programmer could generalize for an specific code.  

The rest of this work is organized as follows: section \ref{sec:motivation} dwells into some of our motivations for proposing this method, in section \ref{sec:related-work} we present the related work, section \ref{sec:instrumentation-tool} presents the instrumentation tool, section \ref{sec:tracing-library} introduces the tracing library for collecting the necessary information for building the execution graph, section \ref{sec:execution-graph} presents what is and how to generate the execution graph from the collected trace, section \ref{sec:graph-analysis} introduces concepts for exploring parallelism in directed graphs and presents an algorithm for exploring parallelism in the execution graph, \ref{sec:results} presents some results obtained from the proposed approach, we conclude on section \ref{sec:conclusions}. 

\section{Motivation}\label{sec:motivation}
The need for porting serial codes into parallel ones creates a natural need for tools capable of aiding and facilitating the job of programmers in the quest of parallelization. There exist two main categories for code analysis methods, namely: static and dynamic analysis. The first one happens at compile time, while the second happens on runtime and requires program execution. Both techniques present drawbacks and strengths, for example there exists important information that will never be available at compile time as well that a program execution might not generate generalizable results.

Examples of dynamic and static analysis tools are Intel® Inspector\cite{intel-inspector-web}, CppDepend \cite{cppdepend-web}, Parallelware Analyzer \cite{parallelware-web} and many others, however almost all code analysis tools are of proprietary nature, thus cannot be used or integrated into open source compiler pipelines like Clang \cite{llvm} or GCC \cite{gcc} due to their closed nature. This creates an important vacuum in the area of code analysis for porting serial codes to parallel programming models, thus the need to create open, modern and extensible tools like what the LLVM project initially created for the compiler community. 

\section{Related Work}\label{sec:related-work}
The concept of visualizing tasks and their dependencies is a well established idea, as it has a straightforward meaning, however it remains an open research area. With papers tackling different aspects of the area, for example in \cite{10.1145/2835238.2835240} they explore how to relate the scheduler and the dependencies and visualize them in an effective way. In \cite{ 8805434} they present an interactive analysis visualization tool for execution graphs, with their goal not being automated analysis rather an interactive one performed by the user and aided by the tool. 

Exploring and analyzing the execution graph of a sequential program for discovering parallelism is an active research area, with new publications published each year. One of the most prominent works in the area is the DiscoPoP profiler first introduced in \cite{6687444} and expanded in \cite{10.1007/978-3-319-27140-8_39, 10.1145/2723772.2723777, tuprints5741, norouzi_ea:ics:2019, https://doi.org/10.1002/cpe.4770, 10.1145/3087556.3087592, 7516000}. DiscoPoP is a tool for discovering parallelism based on the idea of Computational Units (CU)\cite{6687444}, analyzing dependencies between the CU and pointing into likely parallelization opportunities.

The main difference between our approach and DiscoPoP is that we require the user to input the regions to study and then obtain the parallelization opportunities based on abstract approach rather than matching to known patterns \cite{norouzi_ea:ics:2019}.

%% file: tex/2_tool.tex
\ssection{Instrumentalization tool}{instrumentation-tool}
In this section we present a tool using the Clang-LLVM compiler infrastructure \cite{llvm}, responsible for adding instrumentation to an annotated input source code written in C/C++ and producing an instrumented version of the program, ready for trace collection using the trace collection library presented in section \ref{sec:tracing-library}. For a small introduction to Clang-LLVM internals see \cite{aosa}.

The input source code needs to be annotated with pragmas indicating the tracing and tasking regions of interest to be analyzed, the syntax needed for the annotations is presented in Figure \ref{lst:mt:pragma}. Alternatively we provide a C/C++ API for defining the tracing and tasking regions, with its correspondent syntax presented in Figure \ref{lst:mt:api}. We provide this second annotation method because in order to use the first one the user needs to compile a modified version of the Clang compiler. An example of an annotated function is presented in Figure \ref{fig:eg:sw}.

\begin{figure}[h]
    \centering
    \begin{subfigure}[b]{0.5\textwidth}
        \centering
\begin{lstlisting}[frame=tlrb]
#pragma exg trace // Trace block
{
#pragma exg task  // Task block
{
<...>         // Block of code
}
<...>         // Block of code
}
\end{lstlisting}
        \caption{Pragma syntax used to annotate the input source code for defining the tracing and tasking regions.}
        \label{lst:mt:pragma}
    \end{subfigure}%
    ~ 
    \begin{subfigure}[b]{0.5\textwidth}
        \centering
\begin{lstlisting}[frame=tlrb]
mt_btrace(); // Begin trace
<...>        // Block of code
mt_btask();  // Begin task
<...>        // Block of code
mt_etask();  // End task
<...>        // Block of code
<...>        // Block of code
mt_etrace(); // End trace
\end{lstlisting}
        \caption{C/C++ API syntax used to annotate the input source code for defining the tracing and tasking regions.}
        \label{lst:mt:api}
    \end{subfigure}
    \caption{C/C++ syntax for defining the tracing and tasking regions.}
\end{figure}

The instrumentation of the code is performed by an LLVM pass schedule to perform after the LLVM optimization passes. The pass works by inserting into the LLVM-IR, calls to the functions of the tracing library presented in Figure \ref{lst:api:lib}. Specifically it adds calls to the \lstinline{__mt_trace_ir} function every time a \textbf{store} or \textbf{load} instruction is detected, passing the address of the memory position being accessed and an unique constant numeric identifier of the specific instrumentation, to the function. 

Additionally it will also generate unique numeric identifiers for each of the tasks and tracing regions, calling \lstinline{__mt_btrace} when a trace region begins and \lstinline{__mt_etrace} when a trace region ends, \lstinline{__mt_btask} when a task region begins and \lstinline{__mt_etask} when a task region ends.

The pass detects that a region began or ended by either encountering a call to the functions in Figure \ref{lst:mt:api} or encountering a particular metadata node associated to one of the pragmas in \ref{lst:mt:pragma} and generated by the CodeGen phase of the Clang compiler. This later option requires a modified version of the Clang compiler, were the modifications extend the behavior of the preprocessor, parser, semantic analyzer and the code generation phase to accept the pragma as a valid C/C++ construct.

Once the source code is processed by the instrumentalization tool, it needs to be linked against the trace collection library described in section \ref{sec:tracing-library}. The commands needed for obtaining the executable from a source file are presented in Figure \ref{lst:sh:cc}. 
Observe that the code gets compiled with debug symbols as this enables to map the LLVM-IR to the high-level source code. 

\begin{figure}[h]
    \centering
\begin{lstlisting}
clang -I<trace library include path> -g -O1 -emit-llvm -S -o example.bc -Xclang -load -Xclang libinstrumentalization.so example.c
clang example.bc -o example.exe -lmemory_tracer -lstdc++
\end{lstlisting}
    \caption{Clang commands to instrument the source file: \emph{example.c}.}
    \label{lst:sh:cc}
\end{figure}

\ssection{Trace collection library}{tracing-library}
The trace collection library handles metric collection upon execution of the instrumented program. The collection process does not interfere with the normal runtime behavior of the program, it only observes the interactions of the tasks and the instrumented memory positions, records them and finally produces at program exit a binary file \lstinline{trace.out} with the program trace. The public API of the library is presented in Figure \ref{lst:api:lib}.

\begin{figure}[h]
    \centering
\begin{lstlisting}[frame=tlrb]
/// Begin a trace region
/// @param id an unique identifier for the trace region
void __mt_btrace(uint32_t id);
/// End a trace region
void __mt_etrace();
/// Begin a task region
/// @param id an unique identifier for the task region
void __mt_btask(uint32_t id);
/// End a task region
void __mt_etask();
/// Trace a memory access
/// @param address to trace
/// @param id an unique identifier for the instruction accessing the address
void __mt_trace_ir(const void* address, uint32_t id);
\end{lstlisting}
    \caption{API of the tracing library.}
    \label{lst:api:lib}
\end{figure}

Internally the trace collection library works as follows:
\begin{itemize}
    \item Each time that \lstinline{__mt_btrace} gets invoked, create a new trace id.
    \item Every time \lstinline{__mt_btask} is called, create a new execution id such that it is higher than any of the already given and push it into a stack maintaining the current execution id.
    \item When \lstinline{__mt_etask} is called, pop an element from the stack maintaining the current execution id.
    \item Every time \lstinline{__mt_trace_ir} is invoked, retrieve the current execution id and the current trace region id and save the trace id, the execution id, the accessed address, the id of the instruction into the back of list keeping all the traces.
    \item On program exit, save all the lists with the traces into a file.
\end{itemize}

\ssection{Execution graph}{execution-graph}
Once the program trace has been collected it is possible to build an execution graph for the trace, thus we begin by defining the notion of the execution graph.

\begin{definition}
    The execution graph  of a program trace, is a directed graph $G = (V, E)$ in which the vertices $V$ represent task instances and the edges $E$ are execution dependencies between the task instances. A task instance is a task region in the source code together with an execution id. 
\end{definition}

\begin{figure}[h]
    \centering
    \begin{subfigure}[b]{0.5\textwidth}
        \centering
\begin{lstlisting}[frame=tlrb]
void sw(M m, char* s1, char* s2) {
#pragma exg trace
 for(int i = 1; i < M.n(); ++i)
  for(int j = 1; j < M.m(); ++j)
#pragma exg task // Region 1
  {
   int sc = (s1[i-1] = s2[j-1]) ? 
            match : miss;
   m(i, j) = max(m(i, j) + sc,
			     m(i, j - 1) + gap,
			     m(i - 1, j) + gap, 
			     0);
  }
}
\end{lstlisting}
        \caption{Tracing and tasking annotations on a basic version of the Smith-Waterman algorithm.}
        \label{lst:src:sws}
    \end{subfigure}%
    ~ 
    \begin{subfigure}[b]{0.5\textwidth}
        \centering
        \includegraphics[width = \textwidth, height = 2.5in, keepaspectratio]{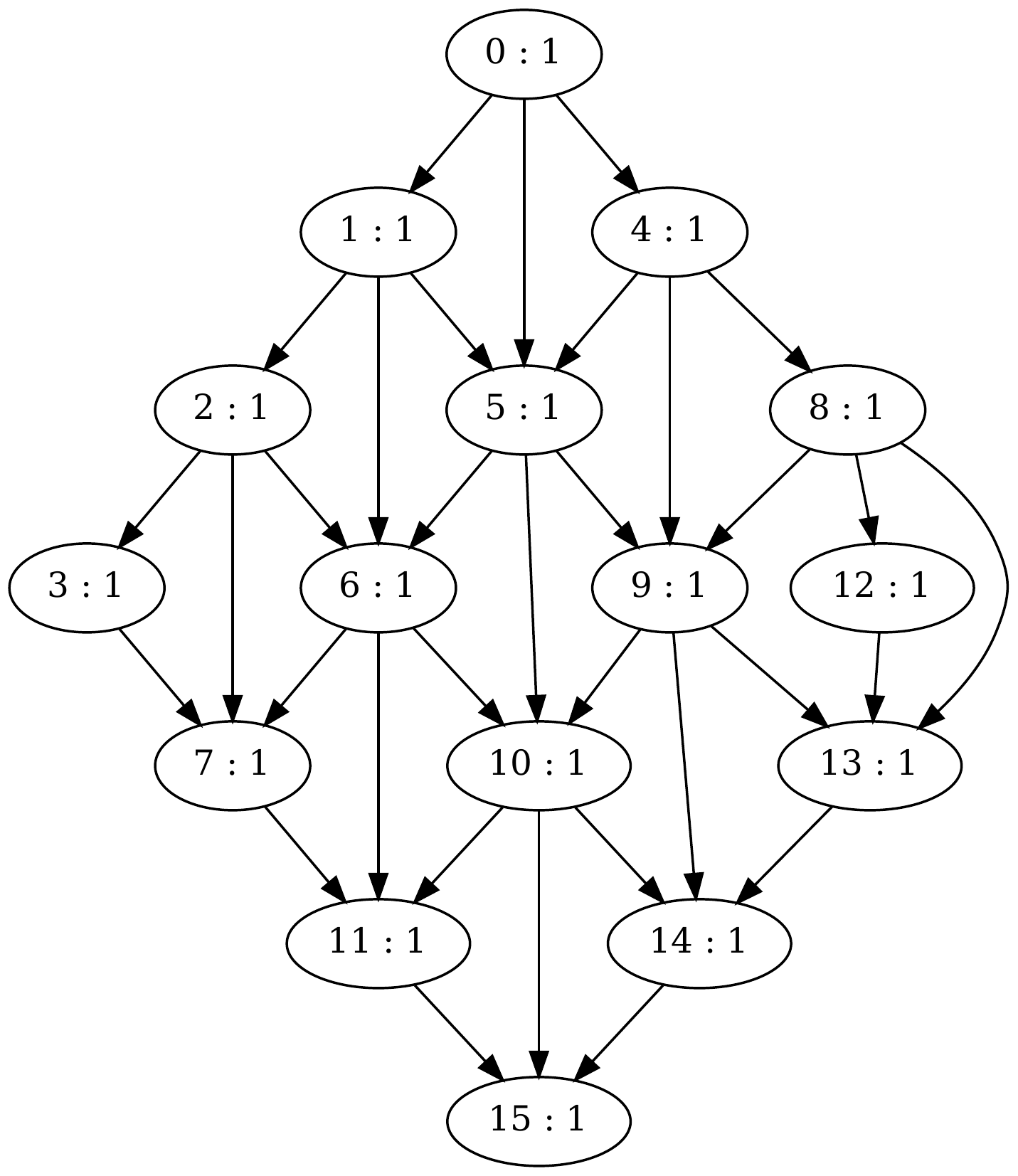}
        \caption{Execution graph obtained by tracing the code presented in LHS (Figure \ref{lst:src:sws}) with sequences of length $4$.}
        \label{fig:eg:sws}
    \end{subfigure}
    \caption{Annotations on a basic version of Smith-Waterman and the generated execution graph produced by those annotations.}
\end{figure}

In particular, we focus this work to the case where the dependencies are data dependencies. An example of  annotated source code using the annotation syntax described in \ref{sec:instrumentation-tool} and the resulting execution graph for some program trace and built with the algorithms presented later in this section, is presented in Figure \ref{fig:eg:sws}. The vertex labels in the execution graph have the following meaning, the first number is the execution id, the second number is the identifier for that region in the source code. The edges represent the data dependencies between the tasks.

The procedure for building the execution graph is presented in Algorithm \ref{alg:ueg}; this algorithm builds the graph one instruction trace at a time. The \textproc{Valid-ID} function simply checks if a task instance is valid. The data structure for holding the directed graph $G$ must be capable of performing the \textproc{Insert-Edge} operation for inserting an edge between two task instances and storing the dependency kind between them, this operation insert the vertices in the graph if they are not already in the graph.

The data structure for holding the address table $AT$ must be capable of performing the operations:
\begin{itemize}
    \item \textproc{Find-Address}: for finding an entry based on a memory address.
    \item \textproc{Update-Entry}: for updating a table entry with new data.
\end{itemize}

\begin{algorithm}[h]
    \begin{tabular}{@{\hspace{0in}}l@{\hspace{0.5em}}r@{\hspace{0.5em}}l}
    \hspace*{\algorithmicindent} \textbf{Inputs}  &  $ptrace$: & program trace\\
    \hspace*{\algorithmicindent} \textbf{Outputs}  &  $G$: & execution graph \\
    \end{tabular}
    \begin{algorithmic}[1]
    \Procedure{Build-EG}{ptrace}
        \State $G \gets $ empty directed graph. \Comment{Execution graph}
        \State $AT \gets $ empty address table.
        \ForAll{trace $ \in $ ptrace} \Comment{Go through the trace list in order}
        \State $a \gets $ trace.address \Comment{Accessed memory position}
        \State $t \gets $ trace.task\_instance \Comment{Task instance accessing $a$}
        \State $rw \gets $ trace.access\_kind \Comment{Either \textbf{read} or \textbf{write}}
        \State $(t', k) \gets $ \Call{Update-Table}{$AT$, $a$, $t$, $rw$}
        \If{\Call{Valid-ID}{$t'$}}
            \State \Call{Add-Dep}{$G$, $(t',\, t)$, $k$}
        \EndIf
        \EndFor
        \State \textbf{return} $G$
    \EndProcedure
    \end{algorithmic}
    \caption[Update Execution Graph.]{Algorithm for building the Execution Graph.}
    \label{alg:ueg}
\end{algorithm}

The procedure for updating the address table is presented in Algorithm \ref{alg:ut}.  The purpose of the address table $AT$ is to store the recent history of every memory address accessed by the program, so that we can determine who was the last task writing to a certain memory position and establish dependencies. There are three kind of dependencies, $\text{RAW}$ read after write, $\text{WAR}$ write after read and $\text{WAW}$ write after write, also known respectively as true dependency, anti-dependency and output dependency \cite{solihin2015fundamentals}. We often will ignore $\text{WAR}$ and $\text{WAW}$ as they can be safely removed by variable renaming or address duplication \cite{solihin2015fundamentals}.

\begin{algorithm}[h]
    \begin{tabular}{@{\hspace{0in}}l@{\hspace{0.5em}}r@{\hspace{0.5em}}l}
    \hspace*{\algorithmicindent} \textbf{Inputs} & $AT$: & the address table \\
    & $a$: & accessed memory address \\
    & $t$: & the task instance accessing $a$ \\
    & $rw$: & one of \textbf{write} or \textbf{read} \\
    \hspace*{\algorithmicindent} \textbf{Output} & $t'$: & the previous task instance that modified the status of $a$ in $AT$ \\
    & $k$: & the kind of dependency generated by the access to $a$
    \end{tabular}
    \begin{algorithmic}[1]
    \Procedure{Update-Table}{$AT,\, a,\, t,\, rw$}
        \State $(t', rw') \gets $ \Call{Find-Address}{$AT$, $a$}
        \State \Call{Update-Entry}{$AT$, $a$, $(t,\, rw)$}
        \If{\Call{Valid-ID}{$t'$}}
            \If{ $rw' = \text{ read}\,\wedge\, rw = \text{ write}$ }
                \State \textbf{return} $(t', \text{WAR})$
            \ElsIf{ $rw' = \text{ write}\,\wedge\, rw = \text{ read}$ }
                \State \textbf{return} $(t', \text{RAW})$
            \ElsIf{ $rw' = \text{ write}\,\wedge\, rw = \text{ write}$ }
                \State \textbf{return} $(t', \text{WAW})$
            \EndIf
        \Else
            \State \textbf{return} $(\text{NULL},\, \text{NULL})$
        \EndIf
    \EndProcedure
    \end{algorithmic}
    \caption[Address table update.]{Algorithm for updating the address table.}
    \label{alg:ut}
\end{algorithm}

Finally we present two procedures for adding dependencies to the execution graph. The first one calls directly the \textproc{Insert-Edge} function and inserts the edge irregardless of anything, while the second, Algorithm \ref{alg:mg2}, prevents the creation of cycles in the graph by inserting an extension dependency \text{EXT}. The second algorithm uses the function \textproc{Renew-ID}, which assigns a new execution id to the task instance $t$ higher than all executions ids in the program trace and the function \textproc{ID} which returns the execution id of a task instance.

\begin{algorithm}[h]
    \begin{tabular}{@{\hspace{0in}}l@{\hspace{0.5em}}r@{\hspace{0.5em}}l}
    \hspace*{\algorithmicindent} \textbf{Inputs} & $G$: & the execution graph \\
    & $t$: & the task instance accessing an address $a$ \\
    & $t'$: & the previous task instance accessing the address $a$ \\
    & $k$: & the kind of dependency between $t'$ and $t$
    \end{tabular}
    \begin{algorithmic}[1]
    \Procedure{Add-Dep-Ext}{$G$, $(t',\, t)$, $k$}
        \If{\Call{ID}{$t'$} $ \geq $ \Call{ID}{$t$} }
            \State $t''\gets $ \Call{Renew-ID}{$t$}
            \State \Call{Insert-Edge}{$G$, $(t, t''),\, \text{EXT}$}
            \State \Call{Insert-Edge}{$G$, $(t', t''),\, k$}
        \Else
            \State \Call{Insert-Edge}{$G$, $(t', t),\, k$}
        \EndIf
    \EndProcedure
    \end{algorithmic}
    \caption[Algorithm for adding dependencies no. 2.]{Algorithm for adding dependencies between tasks using extension dependencies.}
    \label{alg:mg2}
\end{algorithm}

\begin{proposition}
    \begin{enumerate}
        \item If the program has no nested tasks regions, then the execution graph produced by the \textproc{Build-EG} algorithm in conjunction with the \textproc{Insert-Edge} algorithm is a directed acyclic graph, abbreviated DAG, with the execution ids being a topological order.
        \item The execution graph produced by the \textproc{Build-EG} algorithm in conjunction with the \textproc{Add-Dep-Ext} algorithm is a DAG with the execution ids being a topological order.
    \end{enumerate}
    \begin{proof}~
    \begin{claim*}
            Let $G = (V, E)$ be the execution graph and observe that if $\textproc{ID}(t') < \textproc{ID}(t),\,\forall (t', t) \in E$ then $G$ must not contain any cycles and the ids must be a topological order, as a cycle would imply $\textproc{ID}(t') > \textproc{ID}(t)$ for some $(t', t) \in E$, and the ids happen in an increasing order.  
    \end{claim*}
        \begin{sproof}[1]
            Lets consider a task instance $t$ and observe that all the memory accesses that happened in the trace from the moment $t$ began execution, to the moment it finished, must have been associated with the task instance $t$ as there were no other task instances on the stack of the tracing library in section \ref{sec:tracing-library}, and since that is true for any other tasks, we must have that $t$ can only depend on tasks with a lower execution id, thus the generated graph can not have cycles and the ids form an topological order.
        \end{sproof}
        \begin{sproof}[2]
            Let $(t', t) \in E$ and observe that the \textproc{Add-Dep-Ext} algorithm always forces $\textproc{ID}(t') < \textproc{ID}(t)$, thus the claim proofs the result.
        \end{sproof}
    \end{proof}
\end{proposition}

%% file: tex/3_execution.tex
\ssection{Graph Analysis}{graph-analysis}
In this section we explore properties of directed graphs and provide an algorithm for how to parallelize a certain class of DAGs.

Throughout this section we will use the following notation:
\begin{description}[font=\bfseries, leftmargin=2.5cm, style=nextline]
    \item[$G = (V, E)$] A graph $G$, where $V$ is the set of vertices and $E$ the set of edges of the graph.
    \item[$u \Edge v$] A directed edge in between the vertices $u$ and $v$ in a directed graph $G = (V, E)$.
    \item[$u \Path v$] The existence of a directed path between $u$ and $v$ in a directed graph $G = (V, E)$.
    \item[$\Equiv(X)$] The set of all equivalence relations over the elements of a set $X$.
    \item[{$\left[ x \right]_R$}] The equivalence class of $x\in X$ under $R \in \Equiv(X)$ for some set $X$.
    \item[$\abs{X}$] The cardinality of a set $X$.
\end{description}


\tsection{Vertex independence}{ind}
\begin{definition}
    Let $G = (V, E)$ be a directed graph. Two vertices $v, u \in G$ are said to be independent, denoted by $v \perp u$, if $u \not\Path v$ and $v \not\Path u$.
    \begin{itemize}
        \item We define the set $[v]_\perp$ as $\{ u \in V : u \perp v\} \cup \{v\}$.
        \item A set $I \subseteq V$ is said to be independent if $v \perp u,\,\forall v, u \in I$ such that $v \neq u$.
        \item Let $I \subseteq V$ be an independent set, we say that $I$ is maximally independent if $\forall H\subseteq V$ such that $I \subseteq H$ and $H$ independent then $H = I$.
    \end{itemize}
\end{definition}

\begin{proposition}
    Let $G$ be a directed graph and $I \subseteq V$, then:
    \begin{enumerate}
        \item $I \subseteq \bigcap\limits_{x \in I} [x]_\perp$ if and only if $I$ is independent.
        \item $I = \bigcap\limits_{x \in I} [x]_\perp$ if and only if $I$ is maximally independent.
    \end{enumerate}
    \begin{proof}~
    \begin{sproof}[1]
    \begin{sproof}[``$\Rightarrow$'']
        Suppose that $I \subseteq \bigcap\limits_{x \in I} [x]_\perp$ and let  $v, u \in I$. Observe that we must have $u \in [v]_\perp$ and $v \in [u]_\perp$, thus either $v = u$ or $v \perp u$, which in turn proves that $I$ is independent.
    \end{sproof}
    \begin{sproof}[``$\Leftarrow$'']
        Suppose $I \subseteq V$ is independent and let $v\in I$. Observe that for all $u \in I$ we have that $v \in [u]_\perp$, as $v \perp u$ or $v = u$, thus $I \subseteq \bigcap\limits_{u \in I} [u]_\perp$, proving the claim.
    \end{sproof}
    \end{sproof}~
    \begin{sproof}[2]
    \begin{sproof}[``$\Rightarrow$'']
        Suppose $I = \bigcap\limits_{x \in I} [x]_\perp$. We know that $I$ is independent by \emph{1.}. Let $u \in V$ and observe that if $\Tilde{I} = I \cup \{u\}$ is an independent set then we have:
        \begin{equation*}
            \Tilde{I} \subseteq \bigcap\limits_{x \in \Tilde{I}} [x]_\perp  \subseteq \bigcap\limits_{x \in I} [x]_\perp = I,
        \end{equation*}
        thus $\Tilde{I} = I$, hence $I$ is maximally independent.
    \end{sproof}
    \begin{sproof}[``$\Leftarrow$'']
        Suppose $I\subseteq V$ is maximally independent. By \emph{1.} we know that $I \subseteq \bigcap\limits_{v \in I} [v]_\perp$. Let $u\in \bigcap\limits_{v \in I} [v]_\perp$ and observe that $u \perp v,\, \forall v \in I$ or $u \in I$, if $u \perp v,\, \forall v \in I$ then by the maximality of $I$ we must have that $u\in I$, in consequence $I = \bigcap\limits_{v \in I} [v]_\perp$.
    \end{sproof}
    \end{sproof}
    \end{proof}
\end{proposition}


\begin{definition}
    Let $G = (V, E)$ be a finite DAG, with $V = \{v_1,\ldots,v_n\}$.
    \begin{itemize}
        \item We say that $G$ is completely serial if there exists a permutation $i_1,\ldots, i_n$ of the set $\{1,\ldots,n\}$, such that $(v_{i_{k}}, v_{i_{k + 1}}) \in E$ for all $k=1,\ldots, n-1$. See Figure \ref{fig:serial_graph} for an example of a completely serial graph.
            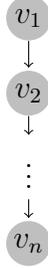
\begin{figure}[h]
                \centering
                \begin{tikzpicture}[shorten >=1pt,->]
                    \tikzstyle{vertex}=[circle,fill=black!25,minimum size=17pt,inner sep=0pt]
                    \node[vertex] (v1) at (0, 3) {$v_1$};
                    \node[vertex] (v2) at (0, 2) {$v_2$};
                    \node (v3) at (0, 1) {$\vdots$};
                    \node[vertex] (v4) at (0, 0) {$v_n$};
                    \draw (v1) -> (v2);
                    \draw (v2) -> (v3);
                    \draw (v3) -> (v4);
                \end{tikzpicture}
                \caption{Completely serial finite DAG.}
                \label{fig:serial_graph}
            \end{figure}
        \item We say that $G$ is completely parallel if $V$ is independent. See Figure \ref{fig:parallel_graph} for an example of a completely parallel graph.
            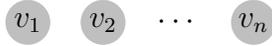
\begin{figure}[h]
                \centering
                \begin{tikzpicture}[shorten >=1pt,->]
                    \tikzstyle{vertex}=[circle,fill=black!25,minimum size=17pt,inner sep=0pt]
                    \node[vertex] (v1) at (0, 0) {$v_1$};
                    \node[vertex] (v2) at (1, 0) {$v_2$};
                    \node (v3) at (2, 0) {$\cdots$};
                    \node[vertex] (v4) at (3, 0) {$v_n$};
                \end{tikzpicture}
                \caption{Completely parallel finite DAG.}
                \label{fig:parallel_graph}
            \end{figure}
    \end{itemize}
\end{definition}

\begin{proposition}\label{prop:completely}
Let $G = (V, E)$ be a finite DAG with $V = \{v_1,\ldots,v_n\}$, then:
    \begin{enumerate}
        \item $G$ is completely serial, if and only if:
        \begin{equation*}
            \abs{[v]_\perp} = 1,\, \forall v\in V.
        \end{equation*}
        \item $G$ is completely parallel, if and only if:
        \begin{equation*}
            \abs{[v]_\perp} = \abs{V},\, \forall v\in V.
        \end{equation*}
    \end{enumerate}
    \begin{proof}~
    \begin{sproof}[1]
    \begin{sproof}[``$\Rightarrow$'']
        Suppose that $G$ is completely serial and let $i_1,\ldots, i_n$ be a permutation of the set $\{1,\ldots,n\}$, such that $(v_{i_{k}}, v_{i_{k + 1}}) \in E$ for all $k=1,\ldots, n-1$. Let $v \in V$ and observe that $\forall u \in V$ we have that either $v \Path u$, $u \Path v$ or $v = u$, as $v = v_{i_j}$ and $u = v_{i_l}$ for some $1\leq j, l\leq n$ and either $l \leq k$ or $k \leq l$.
    \end{sproof}
    \begin{sproof}[``$\Leftarrow$'']
        Suppose that $\abs{[v]_\perp} = 1,\, \forall v\in V$, which translates to $v \not\perp u,\; \forall v, u \in V$ such that $v \neq u$ or equivalently that $\forall v, u \in V$ either $v \Path u$, $u \Path v$ or $v = u$. 
        
        Observe that there exists an unique $v \in V$ such that it has no incoming edges, as $G$ is finite and directed acyclic, uniqueness follows from the fact that if $v' \in V$ is such that it has no incoming edges then either $v' \Path v$ or $v = v'$ thus $v' = v$.
        
        Set $x_1 = v$ and define for $1 < k \leq n$, $x_k \in V$ as the unique vertex such that $x_j \Path x_k$ for $j=1,\ldots,k - 1$ and $u \not\Path x_k,\;\forall u \in V-\{x_1,\ldots,x_{k-1}\}$, existence and uniqueness follows from the acyclicity of the graph and the property $\forall x, u \in V$ either $x \Path u$, $u \Path x$ or $x = u$. Thus $(x_k, x_{k+1}) \in E$ for $k=1,\ldots, n$ and $\{x_1,\ldots,x_{n}\} = V$, hence $G$ is completely serial.
    \end{sproof}
    \end{sproof}
    \vspace{-1cm}
    \begin{sproof}[2]
    \begin{sproof}[``$\Rightarrow$'']
        Suppose that $G$ is completely parallel, then by definition $[v]_\perp = V,\;\forall v\in V$ proving the result. 
    \end{sproof}
    \begin{sproof}[``$\Leftarrow$'']
        Suppose that $\abs{[v]_\perp} = \abs{V},\, \forall v\in V$, then $[v]_\perp = V,\;\forall v\in V $  as there is only one subset of $V$ with $\abs{V}$ elements.
    \end{sproof}
    \end{sproof}
    \vspace{-1cm}
    \end{proof}
\end{proposition}

\begin{definition}
    Let $G = (V, E)$ be a directed graph and $R \in \Equiv(V)$. 
    \begin{itemize}
        \item The quotient graph is defined by:
            \begin{align*}
                G / R &= (V / R, E / R) \\
                V / R &= \{[v]_R : v \in V\} \\
                E / R &= \{([v]_R, [u]_R) : (v, u) \in E \wedge [v]_R \neq [u]_R \}
            \end{align*}
        \item If $G$ is also acyclic we say $R$ is a DAG-preserving if $G / R$ is also a DAG.
    \end{itemize}
\end{definition}

\begin{proposition}
    Let $G$ be a directed graph and $R \in \Equiv(V)$. Let $\hat{u},\, \hat{v} \in V / R\, $ such that $\hat{u}\neq \hat{v}$ then $\hat{u} \perp \hat{v} \implies u \perp v,\; \forall u \in \hat{u},\, v \in \hat{v}$.
    \begin{proof}
        Let $\hat{u},\, \hat{v} \in V / R$ such that $\hat{u}\perp \hat{v}$ and $\hat{u}\neq \hat{v}$, and let $u \in \hat{u}, v \in \hat{v}$. Suppose there is a path between $u$ and $v$ in $G$ given by $u = x_1 \rightarrow \ldots \rightarrow x_k = v$, then $([x_i]_R, [x_{i+1}]_R)_{i=1,\ldots,k - 1}$ is also a path in $G / R$ connecting $\hat{u}$ and $\hat{v}$, which is a contradiction as we assumed $\hat{u}\perp \hat{v}$. Similarly we observe that there is no path between $v$ and $u$. Thus we have proved that $u \perp v$.   
    \end{proof}
\end{proposition}

\begin{definition}
    Let $G = (V, E)$ be a directed graph, we say that $G$ is connected if $G$ is connected as an undirected graph.
\end{definition}

\begin{lemma}
    \label{lem:max_chain}
    Let $G = (V, E)$ be a finite connected DAG and $R \Equiv(V)$ a DAG-preserving relation such that the equivalent classes are independent sets. Suppose $G / R$ is a chain in which every vertex has at most one incoming or one outgoing edge, then there exists a path $\gamma$ of length $\abs{V / R}$ in $G$, furthermore $\gamma$ has maximal path length in $G$.
    \begin{proof}
    Let $\hat{G} = G / R$ be the quotient graph, $\Hat{V} = \{\Hat{V}_1,\ldots, \Hat{V}_n\}$ the set of quotient vertices and $\hat{E} = \{(\Hat{V}_1, \Hat{V}_2), \ldots, (\Hat{V}_{n-1}, \Hat{V}_n)\}$ the set of quotient edges, we suppose that $n > 1$ as the case $n=1$ is immediate. 
    \begin{claim*}
    Let $u_1\rightarrow,\ldots,\rightarrow u_k$ be a path in $G$, then $u_j \in \Hat{V}_{j + m - 1}$ for $j=1,\ldots,k$ and $k \leq n$, where $m$ is such that $[u_1]_R = \Hat{V}_m$.
    \end{claim*}
    Let $i$ be such that $[u_2]_R = \Hat{V}_{i}$ and observe that $i \neq m$ as $u_2 \not\in \Hat{V}_{m}$ because $\Hat{V}_{m}$ is an independent set of vertices in $G$. Next observe that $(\Hat{V}_m, \Hat{V}_{i}) \in \hat{E}$ as $(u_1, u_2)\in E$, thus $i = m + 1$. If we continue this process we obtain $u_j \in \Hat{V}_{j + m - 1}$ for $j=1,\ldots,k$, furthermore since $\hat{G}$ has $n$ vertices we have that $k\leq n$, thus proving the claim.
    
    Let $\Tilde{v}_n$ be a representative of $\Hat{V}_n$ and observe that for all $k < n$ exists $u \in \Hat{V}_k$ such that $u$ and $\Tilde{v}_n$ are connected through a path in $G$. Otherwise $G$ would have at least two connected components, which is a contradiction as we assumed $G$ is connected. 
    
    Let $\gamma$ be a path in $G$ between a member of $\Hat{V}_1$  and $\Tilde{v}_n$. Using the claim we obtain that the length of $\gamma$ in $G$ is $n$, furthermore the claim shows that $\gamma$ has maximal length. 
    \end{proof}
\end{lemma}

\begin{definition}
    Let $G = (V, E)$ be a finite DAG and $R \in \Equiv(V)$.
    \begin{itemize}
        \item We define the execution time of $\Tilde{v} \in V / R$, denoted by $\ExecT(\Tilde{v})$, as:
            \begin{equation*}
                \ExecT(\Tilde{v}) = 
                \begin{aligned}
                    \begin{cases}
                    1 & \text{if } \Tilde{v} \text{ independent as a subset of } V \\
                    \abs{\Tilde{v}} & \text{otherwise}
                    \end{cases}
                \end{aligned}
            \end{equation*}
    
        \item We define the execution time of $G / R$, denoted by $\ExecT(G / R)$, as:
            \begin{equation*}
                \ExecT(G / R) = \sum\limits_{\Tilde{v} \in V / R} \ExecT(\Tilde{v}) 
            \end{equation*}
    \end{itemize}
\end{definition}

\begin{corollary}
    Let $G$ and $R$ be as in Lemma \ref{lem:max_chain} then $\ExecT(G / R)$ is minimal over $\Equiv(V)$.
\begin{proof}
    By Lemma \ref{lem:max_chain}, there is a path $\gamma$ of length $n$. It is easy to observe that $n \leq \ExecT(G / R'),\;\forall R' \in \Equiv(V)$ and that $n = \ExecT(G / R)$, thus proving the result.
\end{proof}
\end{corollary}

\begin{remark}
    Let $G = (V, E)$ be a graph and consider $k$ equivalence relations $R_i \in \Equiv(\hat{V}_i)$ for $i=1,\ldots,k$, where $\hat{V}_1 = V$ and $\hat{V}_{i+1} = \hat{V}_{i} / R_i$ for $i=1\ldots,k$, then $R_k$ induces an equivalence relation in $V$ given by:
    \begin{equation*}
    \begin{split}
        v \;\Tilde{R}_k\; u \iff [[v]_{R_1}\ldots]_{R_k} = [[u]_{R_1}\ldots]_{R_k}.
    \end{split}
    \end{equation*}
    Thus we can also study the graph $G$ by studying quotients of quotients.
\end{remark}

\begin{definition}
    Let $G = (V, E)$ be a graph, the automorphism group of $G$ \cite{Dummit_Foote_2004} is defined by:
    \begin{equation*}
        \Aut(G) = \{ \pi \in \text{BIJ}(V) \mid (v, u) \in E \iff (\pi(v), \pi(u)) \in E \},
    \end{equation*}
    where $\text{BIJ}(V)$ is the set of bijective functions from $V$ to $V$.
\end{definition}

\begin{remark}
    It should be noted that $\Aut(G)$ defines a natural equivalence relation $R$ given by:
    \begin{equation}
        v\, R\, u \iff \exists g \in \Aut(G) \text{ such that } g(v) = u, 
    \end{equation}
\end{remark}

\begin{proposition}
    Let $G = (V, E)$ be a finite DAG then:
    \begin{enumerate}
        \item  All the elements in $V / \Aut(G)$ are independent sets.
        \item  $\Aut(G)$ is DAG-preserving.
    \end{enumerate}
    \begin{proof}~
    \begin{sproof}[1]
        Let $\hat{v} \in V / \Aut(G)$ and let $u, w \in \hat{v}$. By definition there exists an element $\pi \in \Aut(G)$ such that $\pi(u) = w$ and let $\pi^{j+1}(x) = \pi(\pi^j (x))$, with $\pi^0(x) = x$ for $x\in G$. Suppose there exists a path between $u$ and $w$ given by $u = x_1 \rightarrow \ldots \rightarrow x_k = w$.  Observe that:
        \begin{gather*}
            u = x_1 \rightarrow \ldots \rightarrow x_k = w = \pi(x_1) \rightarrow  \ldots \rightarrow \pi(x_k) = \pi^2(x_1)\\
            \pi^2(x_1)  \rightarrow \ldots \rightarrow = \pi^2(x_k) \rightarrow \ldots  \rightarrow  \pi^j(x_1)  \rightarrow \ldots \rightarrow = \pi^j(x_k),
        \end{gather*}
        is also a path on $G$ for all $j \geq 3$, thus $G$ contains an arbitrarily long path which is a contradiction, as $G$ is finite and has no cycles. Hence $\hat{v}$ is an independent set, thus proving 1.
    \end{sproof}
    \begin{sproof}[2]
        Suppose that $\Hat{G} = G / \Aut(G)$ is not a acyclic, let $\Hat{V} = V / \Aut{G}$, then there exists $\hat{x}_1,\ldots,\hat{x}_k \in \Hat{V}$ such that $\hat{x}_1 = \hat{x}_k$ and $(\Hat{x}_j, \Hat{x}_{j+1}) \in E  / \Aut(G),\,\forall j =1, \ldots, k - 1$.
        
        Observe that for each $j=1,\ldots,k-1$ there exists $y^b_j, y^e_j \in V$ such that $[y^b_j]_{\Aut(G)} = \Hat{x}_j,\, [y^e_j]_{\Aut(G)} = \Hat{x}_{j+1}$ and $(y^b_j, y^e_j) \in E$. Let $v_1 \in \Hat{x}_1$, then there exists $\pi \in \Aut(G)$ such that $\pi(y^b_1) = v_1$, set $v_2 = \pi(y^e_1)$ and observe that $(v_1, v_2) \in E$. We can continue this process to obtain $v_{j+1} = \pi_j(y^e_j)$ and $(v_j, v_{j+1})\in E$ with $\pi_j \in \Aut(G)$ such that $v_j = \pi_j(y^b_j)$ for $j = 2,\ldots, k - 1$. If $v_k \not v_1$ then we can find $\sigma \in \Aut(g)$ such that $\sigma(y^b_1) = v_k$ and start the process again, by continuing this scheme either we get a cycle or an infinite path, thus a contradiction, hence $\Hat{G} = G / \Aut(G)$ is acyclic.
    \end{sproof}
    \end{proof}
\end{proposition}

\begin{remark}
    When working with execution graphs we can identify the notion of independence with the notion of parallelism, as independent vertices are tasks instances with no dependencies between each other, thus capable of parallel execution.     
\end{remark}

Finally in Algorithm \ref{alg:sp} we present a procedure for exploring parallelism in codes with highly symmetrical execution graphs.
\begin{algorithm}[h]
    \begin{tabular}{@{\hspace{0in}}l@{\hspace{0.5em}}r@{\hspace{0.5em}}l}
    \hspace*{\algorithmicindent} \textbf{Input}  & $G$: & A DAG \\
    \hspace*{\algorithmicindent} \textbf{Output} & $Q$: & A quotient DAG
    \end{tabular}
    \begin{algorithmic}[1]
    \Procedure{SYM-Explore}{$G$}
        \State $Q \gets G$
        \State $A \gets$ \Call{Automorphism-Group}{G} \Comment{Computes $\Aut(G)$}
        \While{$\abs{A} \neq 1$}
            \State $Q \gets G / A$
            \State $A \gets$ \Call{Automorphism-Group}{$Q$}
        \EndWhile
        \State \textbf{return} $Q$
    \EndProcedure
    \end{algorithmic}
    \caption[computing a symmetrical parallelization.]{Algorithm for computing a symmetrical parallelization.}
    \label{alg:sp}
\end{algorithm}

%% file: tex/4_results.tex
\ssection{Results}{results}
In this section we present the execution graphs generated by our tool, together with the results produced by Algorithm \ref{alg:sp}. We choose $5$ code representatives from 4 of 13 Berkeley's Dwarfs \cite{asanovic2006landscape}, namely, Dense Linear Algebra that entails matrix addition and multiplication, Structured Grids that entails explicit scheme for the one dimensional heat equation, Spectral Methods that entails Fast Fourier transform, and Dynamic Programming that entails Smith-Waterman. 

\begin{figure}[h]
    \centering
    \begin{subfigure}[b]{0.45\textwidth}
    \centering
\begin{lstlisting}[frame=tlrb]
void madd(M &C, const M &A, const M &B) {
#pragma exg trace
  for (size_t i = 0; i < C.n; ++i)
CG: for (size_t j = 0; j < C.m; ++j) 
FG:   C(i, j) = A(i, j) + B(i, j);
}
\end{lstlisting}%
    ~ 
    \caption{C++ code for adding two matrices.}
    \label{lst:src:madd}
\end{subfigure}
    \begin{subfigure}[b]{0.45\textwidth}
        \centering
        \includegraphics[width = \textwidth, height = 2in, keepaspectratio]{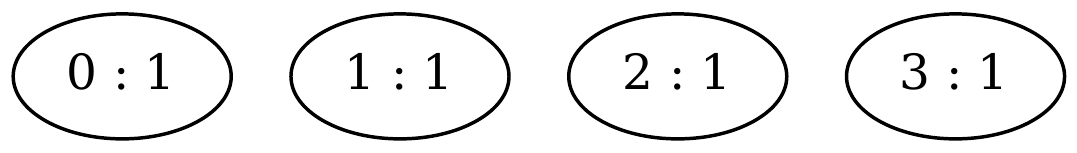}
        \caption{Fine grained execution graph generated by the tool.}
    \label{fig:eg:madd:fg}
    \end{subfigure}\hfil %
    \begin{subfigure}[b]{0.5\textwidth}
        \centering
        \includegraphics[width = 0.5\textwidth, height = 2in, keepaspectratio]{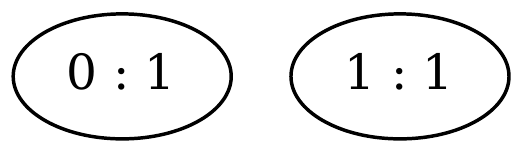}
        \caption{Coarse grained execution graph generated by the tool.}
    \end{subfigure}%
    ~ 
    \begin{subfigure}[b]{0.5\textwidth}
        \centering
        \includegraphics[width = 0.5\textwidth, height = 2in, keepaspectratio]{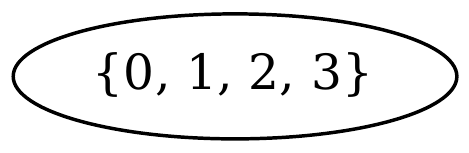}
        \caption{Quotient graph, indicating a parallelization of the program.}
        \label{fig:qu:madd}
    \end{subfigure}
    \caption{Results produced by the tool for the matrix addition function, with matrix sizes of $2\times 2$.}
    \label{fig:eg:madd}
\end{figure}

For each of the test codes our results are presented using 1 figure containing four subfigures. Their organization is as follows: 
\begin{itemize}
    \item The first subFigure is the source code of the function to analyze, in this figure there will be two labels, \lstinline{FG:} and \lstinline{CG:}, indicating placements for the \lstinline{#pragma exg task} construct, providing fine and coarse grain tasking regions. 
        \item The second and third subfigures correspond to the execution graphs generated by the tool for the fine and coarse grain placements respectively, where the first number in the vertex label is the execution id of the task instance and the second one is the region id. 
            \item The fourth and final subfigures is the quotient graph generated by applying Algorithm \ref{alg:sp} to the fine grained execution graph in the second figure, where the vertex labels are the equivalent classes generated by the algorithm, identifying elements by the graph execution ids.
\end{itemize}
\tsection{Matrix addition}{matadd}
The first thing we need to observe from the execution graphs for the matrix addition code, presented in Figure \ref{fig:eg:madd}, is that they are completely parallel. In particular we observe that when we apply  Algorithm \ref{alg:sp}, we obtain a graph with a single node as the quotient graph, meaning that all tasks can be executed at the same time. 
This result can be verified using Proposition \ref{prop:completely}, as the number of independent tasks instances equals the number of vertices in the graph.

\begin{figure}[ht]
    \centering
    \begin{subfigure}[b]{0.45\textwidth}
    \centering
\begin{lstlisting}[frame=tlrb]
void mmult(M &C, const M &A, const M &B) {
#pragma exg trace
 for (size_t i = 0; i < C.n; ++i) 
  for (size_t j = 0; j < C.m; ++j) 
CG: for (size_t k = 0; k < A.m; ++k)
FG:  C(i, j) += A(i, k) * B(k, j);
}
\end{lstlisting}%
    ~ 
    \caption{C++ code for multiplying two matrices.}
    \label{lst:src:mmult}
\end{subfigure}
    \begin{subfigure}[b]{0.45\textwidth}
        \centering
        \includegraphics[width = \textwidth, height = 2in, keepaspectratio]{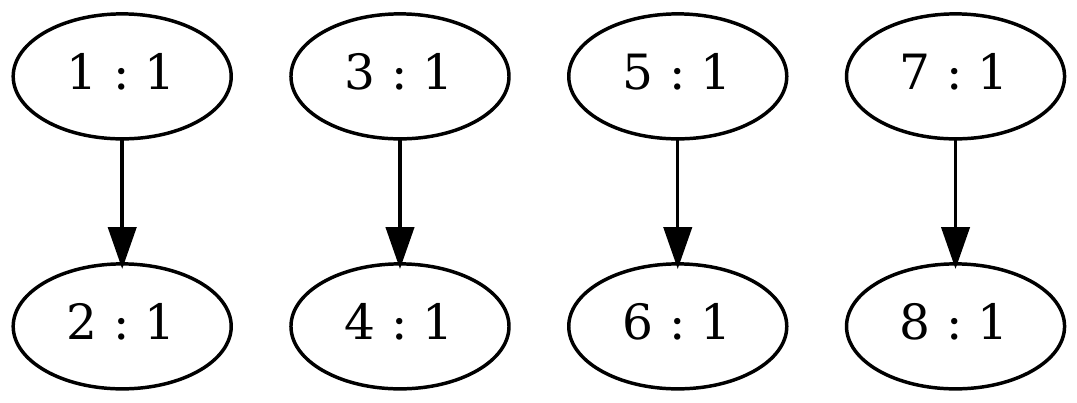}
        \caption{Fine grained execution graph generated by the tool.}
    \label{fig:eg:mmult:fg}
    \end{subfigure}\hfil %
    \begin{subfigure}[b]{0.5\textwidth}
        \centering
        \includegraphics[width = 0.75\textwidth, height = 2in, keepaspectratio]{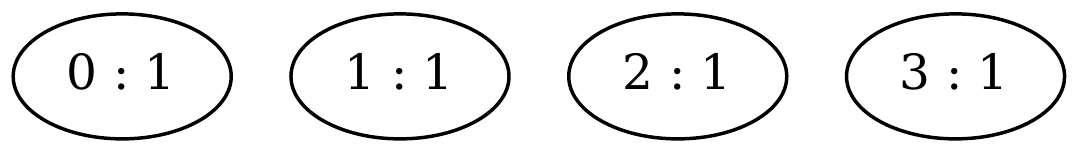}
        \caption{Coarse grained execution graph generated by the tool.}
    \end{subfigure}%
    ~ 
    \begin{subfigure}[b]{0.5\textwidth}
        \centering
        \includegraphics[width = 0.3\textwidth, height = 2in, keepaspectratio]{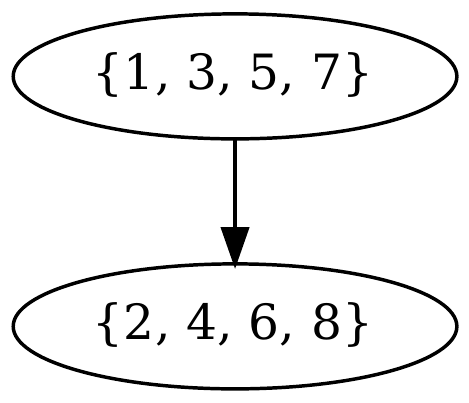}
        \caption{Quotient graph, indicating a parallelization of the program.}
        \label{fig:qu:mmult}
    \end{subfigure}
    \caption{Results produced by the tool for the matrix multiplication function, with matrix sizes of $2\times 2$.}
    \label{fig:eg:mmult}
\end{figure}

\tsection{Matrix multiplication}{matmult}
The execution graphs for the matrix multiplication code, presented in Figure \ref{fig:eg:mmult}, shows that the fined grained tasks suffer from a serialization effect due to the fact that adding the numbers in a single entry depend on other additions to the same entry, we can observe that this effect vanishes in the coarse grained version of execution graph.  With respect to the quotient graph we observe that the Algorithm  \ref{alg:sp} specifies that we need to steps of completely parallel tasks to obtain the final result.

\begin{figure}[h]
    \centering
    \begin{subfigure}[b]{0.45\textwidth}
    \centering
\begin{lstlisting}[frame=tlrb]
void heat(M &u, real h, real k) {
 size_t nt = u.m - 1;
 size_t nx = u.n - 2;
 T r = k / (h * h);
#pragma exg trace
 for (size_t t = 1; t <= nt; ++t) 
CG:for (size_t x = 1; x <= nx; ++x)
FG: u(x, t) = (1 - 2 * r) * 
                 u(x, t - 1) + 
             r * u(x + 1, t - 1) + 
             r * u(x - 1, t - 1);
}
\end{lstlisting}%
    ~ 
    \caption{C++ code for computing the solution to $1D$ heat equation.}
    \label{lst:src:heat}
\end{subfigure}
    \begin{subfigure}[b]{0.45\textwidth}
        \centering
        \includegraphics[width = \textwidth, height = 2in, keepaspectratio]{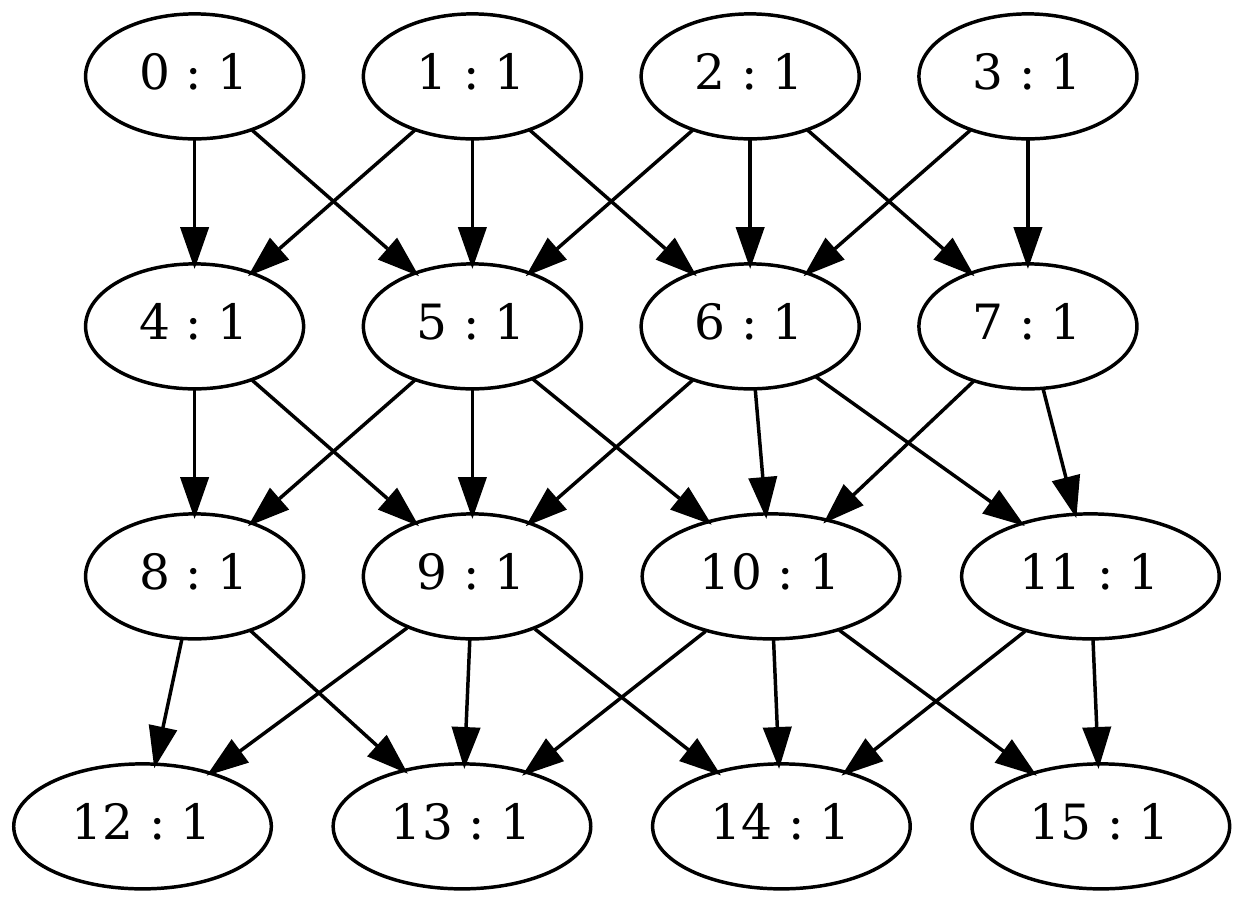}
        \caption{Fine grained execution graph generated by the tool.}
    \label{fig:eg:heat:fg}
    \end{subfigure}\hfil %
    \begin{subfigure}[b]{0.5\textwidth}
        \centering
        \includegraphics[width = 0.4\textwidth, height = 2in, keepaspectratio]{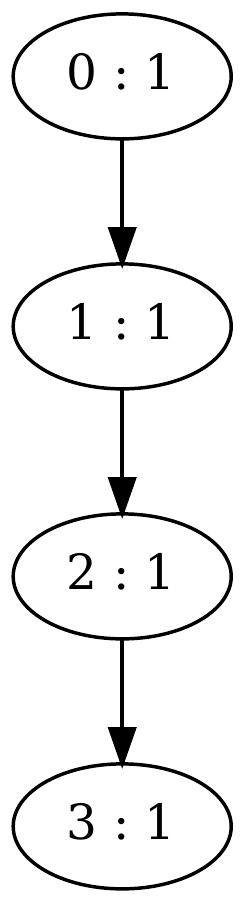}
        \caption{Coarse grained execution graph generated by the tool.}
    \end{subfigure}%
    ~ 
    \begin{subfigure}[b]{0.5\textwidth}
        \centering
        \includegraphics[width = 0.4\textwidth, height = 2in, keepaspectratio]{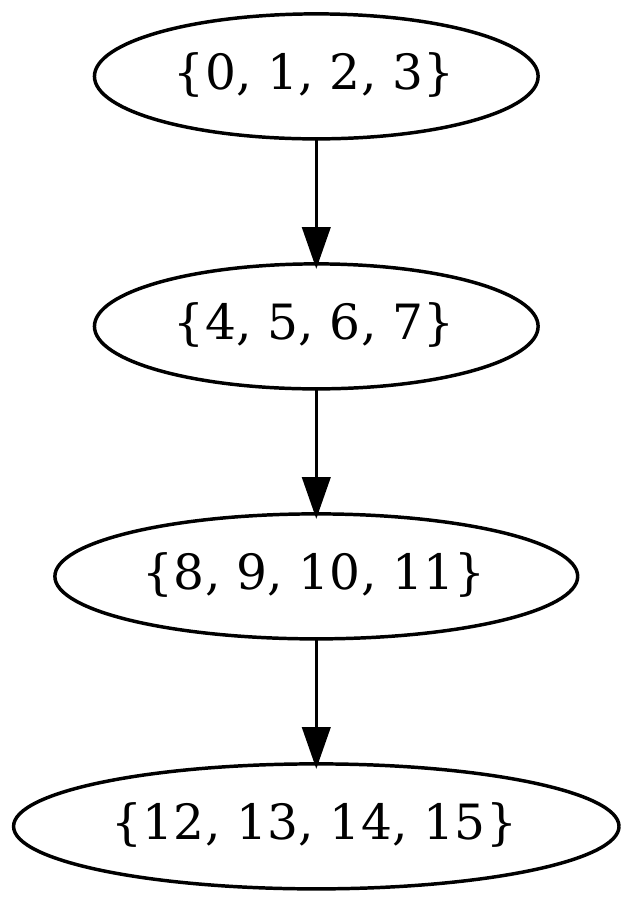}
        \caption{Quotient graph, indicating a parallelization of the program.}
        \label{fig:qu:heat}
    \end{subfigure}
    \caption{Results produced by the tool for the heat equation function, with $4$ grid points and $4$ time steps.}
    \label{fig:eg:heat}
\end{figure}

\tsection{Heat equation}{heat}
In this example we are solving the one dimensional heat equation given by:
\begin{equation*}
    u_t = u_{xx},
\end{equation*}
where $x$ is the spatial dimension and $t$ the temporal one. Specifically the C++ code in Figure \ref{lst:src:heat} corresponds to an explicit finite difference scheme to compute the equation, see \cite{iserles1996first} for more information about the scheme. 

From the execution graph in Figure \ref{fig:eg:heat}, we can see the stencil dependencies through time, with time going from top to bottom. In particular the quotient graph shows the evident, all grid points within a fixed time step can be executed in parallel.

\begin{figure}[H]
    \centering
    \begin{subfigure}[b]{0.45\textwidth}
    \centering
\begin{lstlisting}[frame=tlrb]
void fft(V &X, const V &x) {
 reverse(X, x);
 int l2n = __builtin_ctz(x.size());
#pragma exg trace
 for (int s = 1; s <= l2n; ++s) {
  int m = 1 << s;
  int mh = m >> 1;
  for (int k = 0; k < x.size(); k += m) 
CG:for (int j = 0; j < mh; ++j) FG: {
    T a = (((-2 * j) * M_PI) / m);
    T w = std::exp(a * 1.i);
    T t = w * X[k + j + mh], u = X[k + j];
    X[k + j] = u + t;
    X[k + j + mh] = u - t;
   }
 }
}
\end{lstlisting}%
    ~ 
    \caption{C++ code for the Cooley-Tuckey algorithm.}
    \label{lst:src:fft}
\end{subfigure}
    \begin{subfigure}[b]{0.45\textwidth}
        \centering
        \includegraphics[width = \textwidth, height = 2in, keepaspectratio]{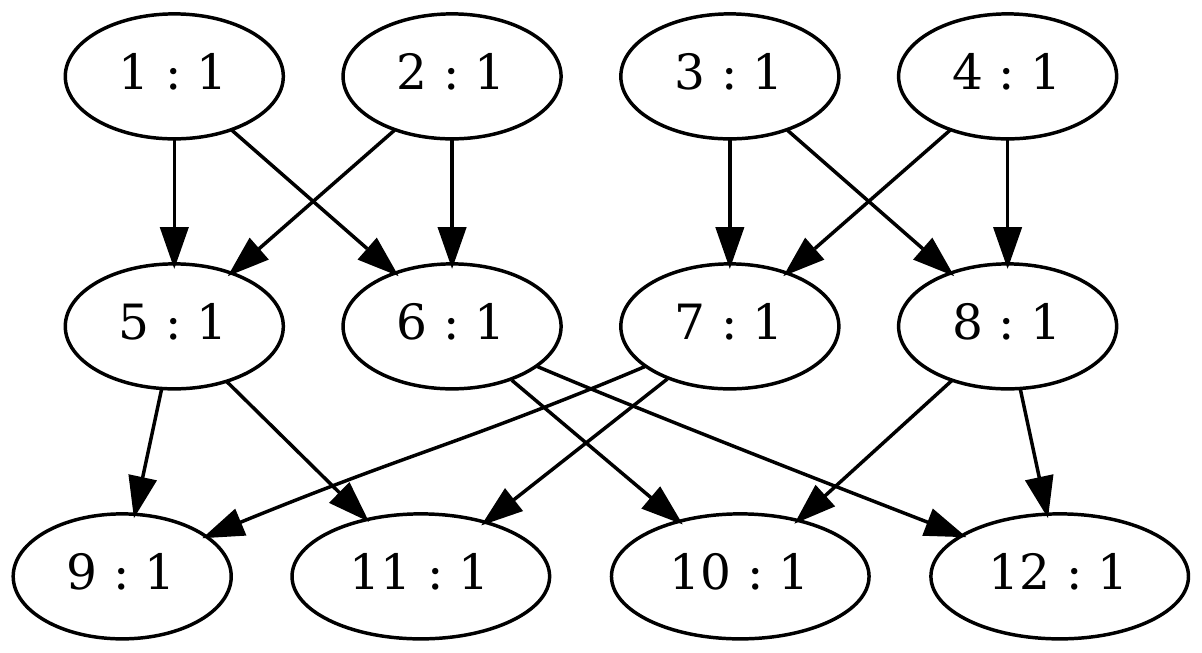}
        \caption{Fine grained execution graph generated by the tool.}
    \label{fig:eg:fft:fg}
    \end{subfigure}\hfil %
    \begin{subfigure}[b]{0.5\textwidth}
        \centering
        \includegraphics[width = 0.4\textwidth, height = 2in, keepaspectratio]{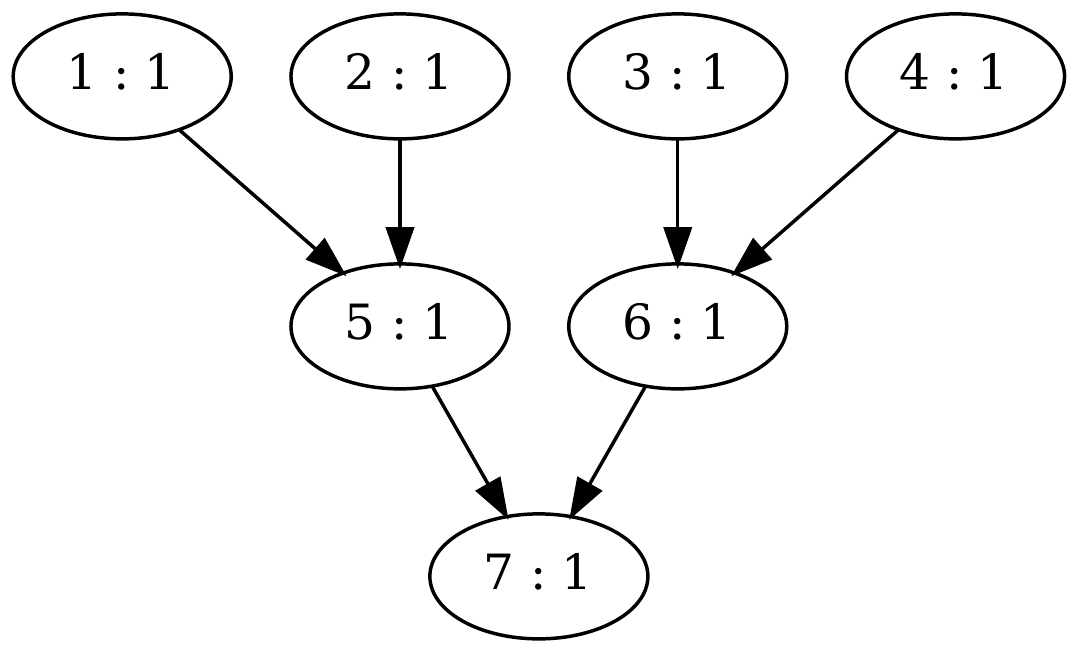}
        \caption{Coarse grained execution graph generated by the tool.}
    \end{subfigure}%
    ~ 
    \begin{subfigure}[b]{0.5\textwidth}
        \centering
        \includegraphics[width = 0.4\textwidth, height = 2in, keepaspectratio]{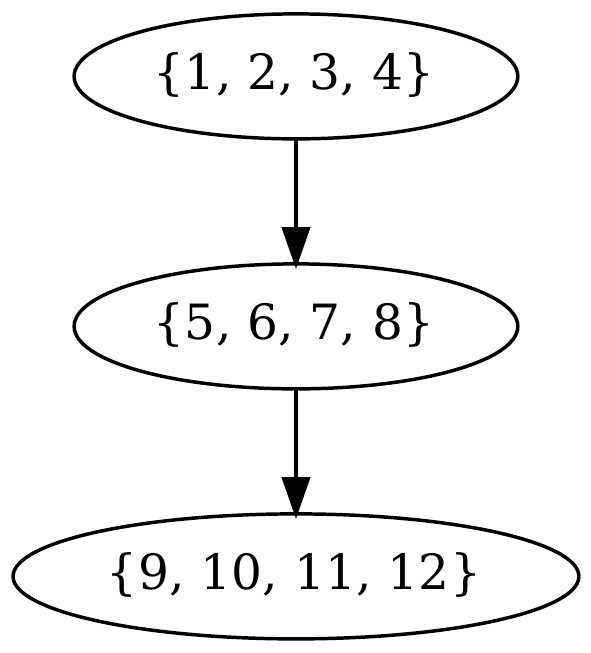}
        \caption{Quotient graph, indicating a parallelization of the program.}
        \label{fig:qu:fft}
    \end{subfigure}
    \caption{Results produced by the tool for fft function, for a vector of length $8$.}
    \label{fig:eg:fft}
\end{figure}

\tsection{Fast Fourier Transform (Cooley-Tuckey)}{fft}
In this example we target the problem of finding the Fast Fourier Transform, for a sequence of $2^n$ numbers using the iterative algorithm presented in \cite{cormen2001introduction}. From the quotient graph in Figure \ref{fig:eg:fft} is possible to observe what it is obvious from the fined grain graph, that the tasks in each of the levels of the tree in the coarse execution graph is independent and can be executed in parallel.

\begin{figure}[h]
    \centering
    \begin{subfigure}[b]{0.45\textwidth}
    \centering
\begin{lstlisting}[frame=tlrb]
void sw(M &m, const S &s1, const S &s2) {
#pragma exg trace
 for(int i = 1; i < M.n(); ++i)
CG:for(int j = 1; j < M.m(); ++j) FG: {
   int sc = (s1[i - 1] == s2[j - 1]) ? 
            match : miss;
   m(i, j) = max(m(i, j) + sc,
                 m(i, j - 1) + gap,
                 m(i - 1, j) + gap, 0);
  }
}
\end{lstlisting}%
    ~ 
    \caption{C++ code for a basic version of the Smith-Waterman algorithm.}
    \label{lst:src:swf}
\end{subfigure}
    \begin{subfigure}[b]{0.45\textwidth}
        \centering
        \includegraphics[width = \textwidth, height = 2in, keepaspectratio]{imgs/sw-fg-4.pdf}
        \caption{Fine grained execution graph generated by the tool.}
    \label{fig:eg:sw:fg}
    \end{subfigure}\hfil %
    \begin{subfigure}[b]{0.5\textwidth}
        \centering
        \includegraphics[width = 0.4\textwidth, height = 2in, keepaspectratio]{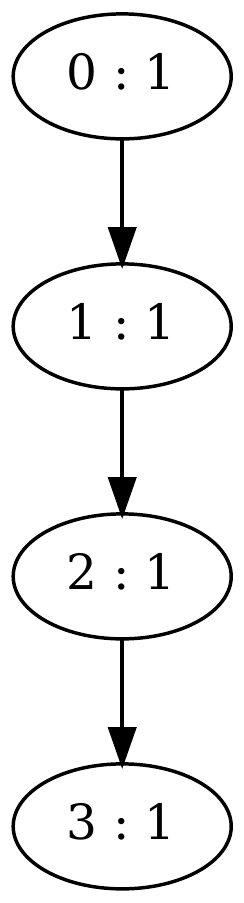}
        \caption{Coarse grained execution graph generated by the tool.}
    \end{subfigure}%
    ~ 
    \begin{subfigure}[b]{0.5\textwidth}
        \centering
        \includegraphics[width = 0.4\textwidth, height = 2in, keepaspectratio]{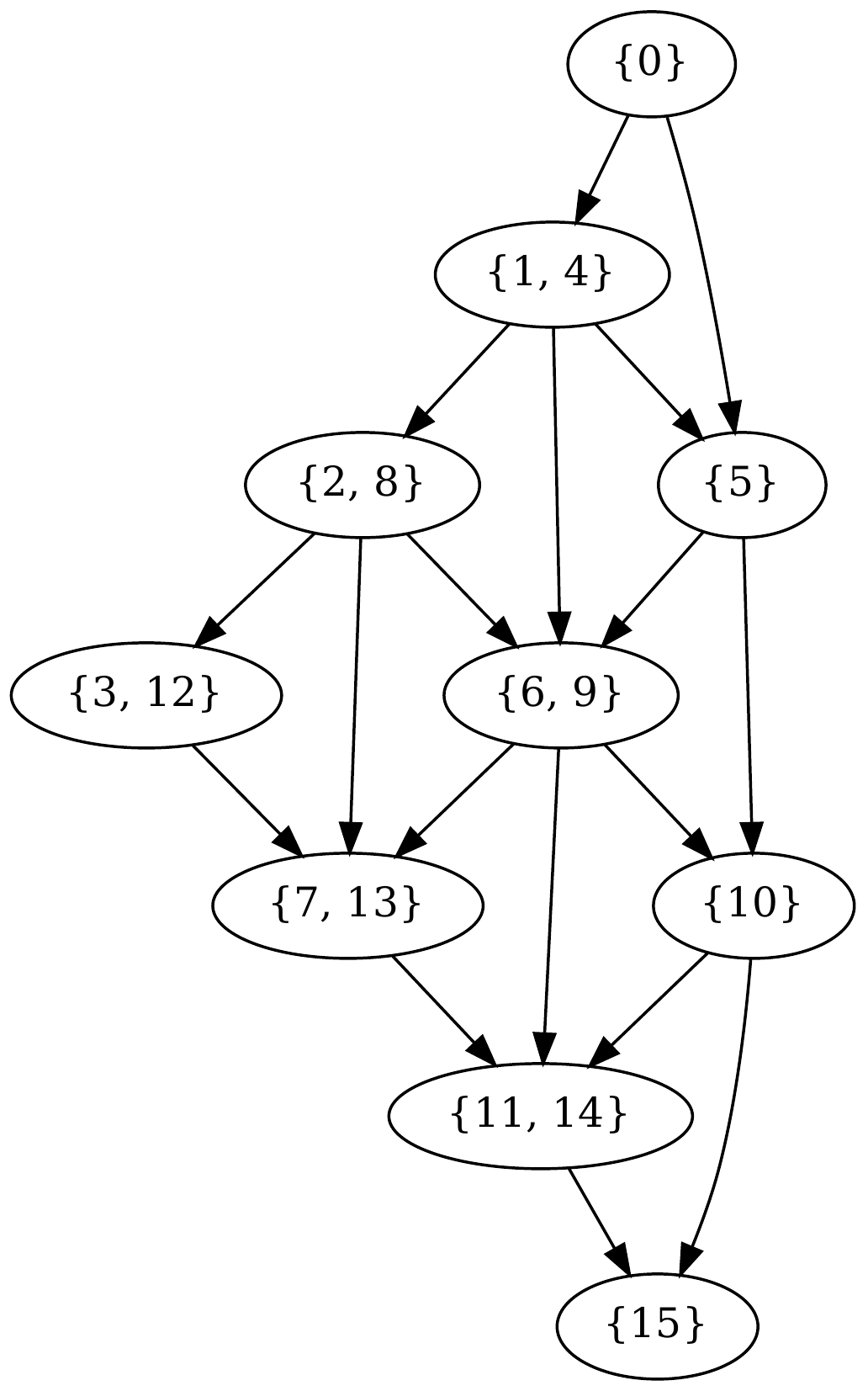}
        \caption{Quotient graph, indicating a parallelization of the program.}
        \label{fig:qu:sw}
    \end{subfigure}
    \caption{Results produced by the tool for the Smith-Waterman function, for two sequences of length 4.}
    \label{fig:eg:sw}
\end{figure}

\tsection{Smith-Waterman}{sw}
In this example we analyze a simple version of the Smith-Waterman algorithm introduced in \cite{SMITH1981195} for computing local sequence alignment. There are two important things to notice from Figure \ref{fig:eg:sw}, that the coarse graph version serializes execution and that the Algorithm \ref{alg:sp} is not able to reduce the graph to a chain, due to the fact that the execution pattern is not that symmetric.

%% file: tex/5_conclusions.tex
\section{Conclusions \& Next Steps}\label{sec:conclusions}

For next steps, we need to explore how to extend the execution graph and the algorithm for building it, to incorporate notions like atomic operations, i.e. operations that could be performed in any order without serialization, thus avoiding task serialization like in the case of the matrix multiply algorithm. Furthermore we need to extend its relationship to the static program structure and eventually to the high-level source code structure, so that the any parallelization detected in the execution graph can be transformed into a potential parallelization of the source code. We also need to extend the analysis techniques and parallelization discovery algorithms to incorporate elements like task duration. 